\documentclass[conference]{IEEEtran}
\IEEEoverridecommandlockouts
\usepackage{cite}
\usepackage{amsmath,amssymb,amsfonts}
\usepackage{algorithmic}
\usepackage{graphicx}
\usepackage{textcomp}
\usepackage{xcolor}
\usepackage{booktabs}
\usepackage{array} % für >{...} in Spaltendefinitionen
\usepackage{comment}

\def\BibTeX{{\rm B\kern-.05em{\sc i\kern-.025em b}\kern-.08em
    T\kern-.1667em\lower.7ex\hbox{E}\kern-.125emX}}
\begin{document}

%\title{Age-Sensitive Gamification in Education:\\ Effects, Age Differences, and Design Principles for Learning Software}
\title{One Size Doesn't Fit All: Age-Aware Gamification Mechanics for Multimedia Learning Environments}

\author{
\IEEEauthorblockN{Sarah Kaißer, Markus Kleffmann \& Kristina Schaaff}
\IEEEauthorblockA{\textit{IU International University of Applied Sciences} \\
Erfurt, Germany \\
\{markus.kleffmann $\mid$ kristina.schaaff\}@iu.org}
}

\maketitle

\begin{abstract}
Gamification is widely used in digital learning. However, most systems neglect age-related differences. This paper investigates how gamification can be designed in an age-aware way to address learners' diverse motivational and cognitive needs. Based on a targeted literature review, we present a mapping of age groups, mechanics, and effects. Furthermore, we derive five design principles for age-specific gamification and identify three technical patterns for implementation in multimedia learning environments. 
The results indicate that gamification is not universally effective, but rather requires a differentiated design to support engagement and inclusivity across the lifespan.
\end{abstract}

\begin{IEEEkeywords}
Gamification, age-aware learning, multimedia learning environments, educational technology, design principles.
\end{IEEEkeywords}

\section{Introduction}\label{secintro}
Multimedia learning technologies play a central role in lifelong learning \cite{becker_zukunftige_2022}. Gamification---the integration of game elements such as points, levels, and progress feedback into learning environments---has been shown to support motivation, engagement, and knowledge integration when aligned with psychological needs \cite{sailer_wirkung_2016, becker_konzept_2022, wagenpfeil_gamification_2025, tondello_framework_2017, sal-de-rellan_gamification_2025}. However, most implementations treat learners as a homogeneous group, overlooking systematic age-related differences in how rewards, progress indicators, and challenges are perceived \cite{noauthor_jahresreport_2024}. This paper addresses this gap by focusing on age-aware gamification.

Psychological research indicates that motivational needs differ substantially across the lifespan \cite{deci_selbstbestimmungstheorie_1993, wagenpfeil_gamification_2025, wesseloh_einsatz_2019, li_gamification_2024}: children benefit from storytelling and multisensory feedback, adolescents from competition and peer recognition, adults from structure and application relevance, and older adults from simplicity and everyday usefulness. However, current approaches rarely translate these insights into concrete design guidelines for multimedia learning environments.

We therefore investigate the following research questions:

\begin{itemize}
    \item \textbf{RQ1:} What are age-related design requirements for gamification?
    \item \textbf{RQ2:} How can gamification mechanics be tailored to different age groups in multimedia learning environments?
\end{itemize}

To answer these questions, we present a theory-led, integrative literature overview and design-oriented synthesis. Our goal is to derive practicable heuristics and design principles for age-adaptive gamification based on Self-Determination Theory (SDT) \cite{ryan_self-determination_2000}, Flow Theory \cite{csikszentmihalyi_beyond_2000}, and Expectancy Theory \cite{vroom_work_1964}.
Our contributions are as follows: 
\begin{itemize}
    \item We introduce an \textit{Age} $\rightarrow$ \textit{Mechanics} $\rightarrow$ \textit{Effects} mapping that links gamification elements to motivational needs across age groups.
    \item We derive five design principles that capture age-specific requirements for gamified educational software.
    \item We outline three technical patterns for integrating age-specific gamification into multimedia learning environments.
\end{itemize}

Together, these contributions provide a blueprint for replacing one-size-fits-all gamification with developmentally informed, multimedia-ready solutions.

%This paper is organized as follows: In Section~\ref{secTheoreticalBackground}, we review the theoretical background for our analysis. In Section~\ref{secMethodology} we describe the applied methodology. In Section~\ref{secResultsAgeRelatedRequirements}, we present the results in terms of age-related requirements and effects. In Section~\ref{secResultsDesignPrinciplesAndPatterns}, we derive design principles and discuss technical implementation patterns. In Section \ref{secDiscussion}, we discuss constraints and ethics. In Section \ref{secConclusionAndFutureWork}, we conclude the paper and outline directions for future work.

This paper is organized as follows: We begin by reviewing the theoretical background (Section~\ref{secTheoreticalBackground}) and then outline our methodology (Section~\ref{secMethodology}). Next, we report age-related requirements and effects (Section~\ref{secResultsAgeRelatedRequirements}) before distilling design principles and technical implementation patterns (Section~\ref{secResultsDesignPrinciplesAndPatterns}). We then examine constraints and ethical considerations (Section~\ref{secDiscussion}) and close with conclusions and directions for future work (Section~\ref{secConclusionAndFutureWork}).

\section{Theoretical Background}\label{secTheoreticalBackground}
In the following, we summarize the conceptual and theoretical foundations of gamification in multimedia learning. We outline core mechanics and their technical realization, distinguish gamification from adjacent paradigms, and review motivational theories.

\subsection{Gamification Concepts and Core Mechanics}
Gamification refers to the integration of game mechanics--such as points, badges, levels, progress indicators, leaderboards, and quests--into non-game contexts to enhance motivation, engagement, and persistence \cite{becker_konzept_2022,becker_entwicklung_2022}.

In educational contexts, gamification seeks to make learning more interactive, immersive, and rewarding by providing learners with structured incentives that increase their willingness to engage with the material and foster extrinsic and intrinsic motivation \cite{deci_selbstbestimmungstheorie_1993}. Such mechanics also promote competition and cooperation \cite{deterding_game_2011, hamari_does_2014, zawacki-richter_synergies_2023}.

Clear challenges, immediate feedback, and social mechanisms are among the most reliable success factors that promote deep concentration and flow \cite{becker_positive_2024,hamari_does_2014}. It has been shown that personalization increases relevance and commitment to the material while storytelling can create emotional involvement~\cite{tondello_framework_2017}. Overall, gamification can raise engagement and strengthen knowledge integration when embedded appropriately \cite{wagenpfeil_gamification_2025}.

\subsection{Adjacent Paradigms}
Gamification is related to, but distinct from, paradigms such as \emph{serious games} and \emph{game-based learning}. Serious games are full-fledged game environments with embedded learning objectives~\cite{zawacki-richter_serious_2023}, and game-based learning integrates complete games into curricula \cite{becker_konzept_2022}. Gamification selectively integrates game elements into existing systems \cite{deterding_game_2011}. In contrast, classic learning software remains more linear and instruction-oriented~\cite{baumgartner_didaktische_2002, bock_digital_2023}. This makes gamification particularly suitable for multimedia systems where core learning content must remain central and not be overshadowed by game narratives.

\subsection{Technical Realization}
From a technical perspective, gamification requires rule and event engines that can trigger feedback based on learner interactions in real time, and coordinate rewards and progression~\cite{pedreira_architecture_2020}. Learning analytics can contribute to personalization by capturing data and shaping adaptive learning paths \cite{wagenpfeil_gamification_2025}.

Implementations typically rely on server-side logic, low-latency communication (e.g., WebSockets), and Application Programming Interfaces (APIs) that connect gamification modules to learning platforms such as Moodle, Canvas, or custom multimedia systems. In addition, learning analytics pipelines can provide the necessary data to personalize challenges, dynamically adjust difficulty, or adapt rewards to learner progress. Increasingly, gamification is combined with artificial intelligence (AI) to enable AI-driven personalization, virtual and augmented reality environments, and mobile learning apps, which expands the spectrum of interaction modalities but also raises challenges for scalability, accessibility, and privacy.

\subsection{Motivation Theories as Analytical Lens}
The effectiveness of gamification has been widely explained through established theories of human motivation:

SDT \cite{ryan_self-determination_2000} distinguishes three psychological needs: \textit{autonomy}, \textit{competence}, and \textit{relatedness}. Game mechanics can fulfill these needs through customization and choice (autonomy), clear feedback and progress indicators (competence), and social features such as collaboration or competition (relatedness). Age groups differ substantially in how strongly they seek each of these needs, making SDT a powerful tool for age-aware gamification.

Flow Theory \cite{csikszentmihalyi_beyond_2000} describes the state of deep engagement that arises when challenge and skill are in balance. Gamification elements such as adaptive difficulty, progressive quests, or real-time feedback can support flow. However, the conditions under which flow emerges differ by developmental stage. For example, children may require scaffolding to avoid frustration, while adults benefit from professional relevance.

Expectancy Theory \cite{vroom_work_1964} explains motivation as a function of three factors: \textit{expectancy} (belief that effort leads to performance), \textit{instrumentality} (belief that performance leads to outcomes), and \textit{valence} (value of these outcomes). Gamification can increase expectancy through transparent rules, strengthen instrumentality through visible cause–and–effect relationships (e.g., points for actions), and enhance valence through meaningful rewards. With age, the relative weight of these factors shifts: children respond strongly to immediate, tangible valence, while adults are more motivated by long-term, competence-oriented outcomes.

\subsection{Implications for Age-Aware Gamification}
Taken together, these perspectives highlight why a one-size-fits-all approach to gamification is insufficient. Psychological needs (SDT), conditions for deep engagement (Flow), and motivational trade-offs (Expectancy Theory) all vary systematically with age. At the same time, technical infrastructures for gamification---rule engines, multimedia interfaces, analytics pipelines---offer mechanisms to implement adaptive, age-specific feedback loops. This dual perspective, combining psychology and technology, provides the foundation for the research questions addressed in this paper.

\section{Methodology}\label{secMethodology}
Our study is based on a targeted literature review combined with a theory-driven synthesis. The aim was to identify age-related requirements for gamification and to derive actionable design principles for multimedia learning environments.

\subsection{Literature Search and Selection}
We used a targeted literature search across multiple databases with a purposeful selection focused on educational context and age-related aspects. Search terms included combinations of \textit{gamification}, \textit{game design}, \textit{game mechanics}, \textit{framework}, \textit{education}, \textit{learning}, \textit{age}, \textit{children}, \textit{adolescent}, \textit{adult}, \textit{self-determination}, \textit{flow}, \textit{expectancy}, \textit{motivation}, \textit{success}, \textit{intrinsic}, \textit{extrinsic}, \textit{feedback}, \textit{reward}, \textit{progress}, \textit{psychological}, \textit{competition}, \textit{collaboration}, \textit{cognitive}, \textit{data}, and \textit{multimedia}.

We screened titles and abstracts against the following inclusion criteria: empirical or conceptual focus on gamification in learning/educational contexts, explicit consideration of age-related factors, learner groups, or developmental stages, publication in peer-reviewed venues (journals, conferences, edited volumes), English or German language, and 2013–2025 publication window (older sources only when providing foundational theories). Exclusion criteria were purely technical papers without educational focus, studies on games without gamification, and non-peer-reviewed sources.

Our overview is narrative and theory-led by SDT, Flow, and Expectancy Theory with a design-oriented synthesis. We report representative findings to derive age-aware heuristics and design principles.

\subsection{Age Group Classification \& Analytical Approach}
To analyze requirements systematically, we categorized the studies into five age groups reflecting common stages of cognitive, social, and educational development which are shown in Table \ref{tab:age_categories}.

\begin{table}[h]
\caption{Learning stages and key characteristics across age groups}
\label{tab:age_categories}
\centering    
\begin{tabular}{@{}%
  >{\raggedright\arraybackslash}p{0.18\linewidth}%
  >{\raggedright\arraybackslash}p{0.23\linewidth}%
  >{\raggedright\arraybackslash}p{0.48\linewidth}@{}}
\toprule
\textbf{Age Group} & \textbf{Stage} & \textbf{Characteristics} \\ \midrule
Children (0--12) & Primary education & Basic literacy and numeracy; high receptivity to play and narrative \\ 
Adolescents (13--17) & Secondary education & Strong orientation toward peers; heightened sensitivity to competition and recognition \\ 
Young Adults (18--25) & Higher education / early career & Increasing autonomy; preference for competence-building and long-term goals \\ 
Adults (26--59) & Professional training & Focus on applicability, efficiency, and structured progress \\ 
Older Adults (60+) & Lifelong learning & Emphasis on simplicity, everyday relevance, and accessibility \\ \bottomrule
\end{tabular}
\end{table}

This classification guided the synthesis of findings, ensuring that gamification requirements were compared and interpreted within coherent developmental categories. While age boundaries are necessarily approximate, participants were categorized into age groups based on international definitions and prior psychological research \cite{unicef_children_1989, who_adolescents_2022, nap_youngadults_2015, who_adults_2022, who_olderadults_2015}. These ranges were chosen to create mutually exclusive and developmentally coherent categories.

%\subsection{Analytical Approach}
The included studies were coded according to the five previously introduced age groups, the gamification mechanics applied, and the reported motivational or learning effects.

To interpret findings systematically, we applied three complementary theoretical approaches: SDT, Flow Theory, and Expectancy Theory. 
This allowed us to connect empirical evidence with psychological models of motivation and to derive requirements that vary across age groups.

\subsection{Synthesis into Design Guidelines}
The extracted requirements were consolidated into an \textit{Age} $\rightarrow$ \textit{Mechanics} $\rightarrow$ \textit{Effects} mapping (Section \ref{secResultsAgeRelatedRequirements}), which served as a basis for the design principles (Section \ref{subsecDesignPrinciples}). In a final step, we derive technical patterns (Section \ref{subsecTechnicalPatterns}) that demonstrate how these principles can be implemented in multimedia learning environments.

\section{Age-Related Effects of Gamification}\label{secResultsAgeRelatedRequirements}
Gamified environments show clear positive effects on motivation, engagement, and learning behavior---with age-specific nuances. Thus, effective gamification differs systematically across developmental stages, with distinct motivators, sensitivities, and challenges for each group \cite{deci_selbstbestimmungstheorie_1993, wagenpfeil_gamification_2025, wesseloh_einsatz_2019}. For example, children and adolescents tend to interact more intuitively with gameful elements, whereas adults are more selective and emphasize relevance \cite{wesseloh_einsatz_2019}. 

To answer RQ1, in this section we analyze age-related requirements for gamification in educational contexts. The synthesis draws on prior literature and empirical findings to identify how motivational, cognitive, and technical needs vary across developmental stages. To provide a comprehensive perspective, we discuss the positive effects of age-appropriate gamification as well as the negative consequences of misapplied mechanics. The analysis is structured by age group and concludes with a summary mapping of requirements, mechanics, and expected effects.

\subsection{Children}
Children are characterized by curiosity, a strong need for autonomy and competence, and high affinity for technology \cite{noauthor_kim-studie_2024}. They particularly benefit from visual, auditory, and haptic cues. Therefore, multisensory environments that combine listening, seeing, and touching facilitate processing and long-term retention \cite{becker_entwicklung_2022}. Popular children's products such as ABCmouse and Endless Alphabet illustrate how animations, puzzles, and stories enable age-appropriate learning \cite{wagenpfeil_gamification_2025, bodynek_gamification_2023}.

Children generally need clear, timely, and visual feedback, as abstract evaluation is harder to grasp at this age \cite{bodynek_gamification_2023, becker_gamification_2022, li_gamification_2024}. Storytelling is essential because narratives emotionally anchor content and create meaningful context \cite{bodynek_gamification_2023}, while movement and discovery games, e.g., escape-game concepts, can foster attention and flow by involving activity and self-efficacy \cite{hes_innovative_2024}. Visible progress and auditory signals can further strengthen emotional engagement \cite{deci_selbstbestimmungstheorie_1993, galliker_kompendium_2015, bodynek_gamification_2023}.

Children’s limited attention spans make them vulnerable to overload when content is unclear or inappropriate for their age. Complex mechanics, unclear feedback, and excessive visual/auditory stimuli (e.g., flashing rewards, constant animations) cause sensory overload, hinder processing, and, according to Cognitive Load Theory, can undermine motivation, especially for younger children in digital environments \cite{plass_foundations_2015}.

%In summary, effective gamification for children leverages multisensory environments that combine visual, auditory, and haptic cues to enhance comprehension and retention. Storytelling, interactive games, and visible feedback further foster motivation, engagement, and self-efficacy. Short and achievable steps are necessary to prevent overload.

\subsection{Adolescents}
For adolescents, clear goals, real-time feedback, and dynamic difficulty adjustments are central \cite{becker_positive_2024, deci_selbstbestimmungstheorie_1993}. Recent studies confirm significant effects on motivation and performance if gamification is applied adequately \cite{sal-de-rellan_gamification_2025}. 

Adolescents value self-determination, competition, cognitive challenge, and socially visible progress \cite{li_gamification_2024}. Systems should provide authentic content and options for co-creation~\cite{deci_selbstbestimmungstheorie_1993}. Feedback should be informative and action-oriented to support orientation and competence \cite{wesseloh_einsatz_2019}.

Since social interaction plays a central role for adolescents, gamified systems should support exchange and cooperation~\cite{wagenpfeil_gamification_2025}. The need for achievement, comparison, and self-efficacy is pronounced. Thus, competition through leaderboards and public progress displays can greatly increase motivation if carefully designed \cite{hanus_assessing_2015}. However, opportunities for collaboration are also important to counterbalance pressure. Thus, clear rules, task structures, and feedback are necessary to create transparency and support self-regulation \cite{wesseloh_einsatz_2019}.

Unlike children, adolescents benefit from more complex mechanics. Quests and group statistics provide structure and peer recognition, thereby supporting both competition and collaboration \cite{wagenpfeil_gamification_2025}.

Gamification can also negatively affect adolescents. Performance-focused leaderboards may induce stress and frustration \cite{wagenpfeil_gamification_2025}. For some, it reduces intrinsic motivation and heightens social comparison, pressure, and dissatisfaction \cite{hanus_assessing_2015}. Constantly visible failures and reward systems perceived as controlling undermine autonomy and self-efficacy, particularly discouraging lower-performing learners.

%In summary, effective gamification for adolescents emphasizes clear goals, real-time feedback, and dynamic challenges to enhance motivation and performance. Social interaction, competition, and co-creation are key drivers, while quests and collaborative features balance pressure and foster both achievement and peer recognition.

\subsection{Young Adults}
Young adults primarily use gamification in higher education, vocational training, or professional up-skilling \cite{faltermaier_entwicklungspsychologie_2014}. They value structured goals, self-determination, collaboration, progress visualization, and personalized feedback \cite{wagenpfeil_gamification_2025, sailer_wirkung_2016}. Quests can foster collective action and engagement \cite{becker_konzept_2022}, rankings support competence experiences \cite{wesseloh_einsatz_2019}, and collaborative learning opens up intercultural perspectives \cite{zawacki-richter_international_2023}. 

Extrinsic rewards lose importance, while functional, uncluttered interfaces gain value \cite{sailer_wirkung_2016}. Comparative features provide orientation rather than rivalry and can support cooperation~\cite{sailer_wirkung_2016}. Motivation is often purpose-oriented and tied to perceived usefulness \cite{galliker_kompendium_2015}. Autonomy is a central need for this group, and informative feedback and choices are vital for competence and responsibility\cite{wesseloh_einsatz_2019}. 

Among young adults, limitations of gamification arise primarily when the playful elements do not align with personal goals and are not compatible with the logic of learning. Unclear or trivial rewards---especially points/badges detached from real learning---feel superficial, devalue achievement, and can undermine self-drive \cite{wagenpfeil_gamification_2025}. Benefits are uneven: disheartened or underachieving students often engage less and may need targeted support \cite{barata_gamification_2015}. Overemphasis on performance or visible failures can lead to frustration, social withdrawal, and demotivation.

%In summary, effective gamification for young adults focuses on structured goals, autonomy, collaboration, and personalized feedback to support purposeful learning. Clear interfaces, quests, and comparative features enhance competence and engagement, while autonomy and informative feedback strengthen responsibility and self-determination.

\subsection{Adults}
Adults often use gamified systems in continuing education or corporate learning \cite{becker_einleitung_2022}. Because they often must coordinate work, family, and learning, flexible, everyday-integrated offerings are especially important \cite{widyakusuma_future_2024}.

For adults, gamification is effective when functionally integrated: storytelling, real-time feedback, and modular content increase engagement when directly related to professional practice \cite{beil_interface_2018}. Studies report consistently medium to high effect on engagement when gamification is adequately applied \cite{looyestyn_does_2017}.

Adults value autonomy-supportive structure and application-oriented tasks \cite{li_gamification_2024}. Motivation increases when learning processes support autonomy, relatedness, and competence \cite{deci_selbstbestimmungstheorie_1993}. In Expectancy Theory terms \cite{vroom_work_1964}, attainable goals with visible utility are particularly effective \cite{galliker_kompendium_2015}.

Adults prefer clear structures, differentiated feedback, and purposeful interfaces that can transfer to real application scenarios \cite{beil_interface_2018}. Overly playful designs are viewed critically---learning environments should feel like purposeful tools.

Adults are sensitive to excessive visuals and pop-ups, and competition-oriented leaderboards are often seen as disruptive or unprofessional \cite{wagenpfeil_gamification_2025}. Incoherent gamification, such as complex interactions, repeated non-disappearing animations, or overly playful/unstructured elements, distracts from goals, raises cognitive load, and creates misaligned incentives.

%In summary, effective gamification for adults emphasizes flexibility, autonomy-supportive structures, and application-oriented tasks that integrate seamlessly into daily and professional contexts. Storytelling, real-time feedback, and purposeful interfaces enhance engagement, while attainable goals and clear structures strengthen motivation and transfer to practice.

\subsection{Older Adults}
Older adults are increasingly in the focus of learning research due to demographic change and the importance of lifelong learning \cite{becker_einleitung_2022}. Age-related limitations, such as reduced memory or motor skills, require simple and accessible designs \cite{fietkau_quests_2019}. Studies suggest that gameful training can improve cognitive flexibility and control when tasks are introduced gradually \cite{basak_can_2008}. Gradual task introduction and adjustable levels of difficulty also support competence and flow \cite{ryan_self-determination_2000, wesseloh_einsatz_2019}.

Older adults greatly benefit from clear orientation (e.g., in the form of progress indicators and attainable goals) and everyday applicability \cite{li_gamification_2024}. Motivation derives from personal relevance, social participation, and achievable goals \cite{fietkau_quests_2019}. Easily understandable feedback further fosters motivation, while small successes strengthen self-efficacy and persistence \cite{technische_universitat_berlin_fakultat_fur_verkehrs-_und_maschinensysteme_gate_2015}. Accessible interfaces such as high contrast, audio support, or read-aloud functions are particularly important for this age group and further increase engagement \cite{technische_universitat_berlin_fakultat_fur_verkehrs-_und_maschinensysteme_gate_2015}.

Older adults need highly usable, low-complexity designs. Unclear navigation, tiny controls, and cluttered interfaces are major barriers; tools not adapted to sensory limits or limited digital experience are problematic \cite{technische_universitat_berlin_fakultat_fur_verkehrs-_und_maschinensysteme_gate_2015}. Poor learning outcomes often result less from the content itself than from the cognitively overwhelming access to it. Overstimulation, complex interactions, and insufficient support demotivate and hinder participation \cite{stieglitz_gamification_2015}. Weak structure, poor accessibility, and complexity trigger overload and reduce confidence.

%In summary, effective gamification for older adults relies on simple, accessible designs, gradual task introduction, and clear orientation to accommodate age-related limitations. Everyday relevance, achievable goals, and supportive feedback foster motivation, while accessible interfaces and small successes strengthen engagement, self-efficacy, and persistence.

\subsection{Age-Related Motivators for Gamification}
As shown in the previous sections, no universal set of mechanics is equally effective across all age groups. Requirements vary systematically with developmental stage, motivational needs, and media experience. Table \ref{tab:age_mechanics_effects} consolidates these findings into a mapping that answers RQ1. It also underpins the design and implementation guidance in Section \ref{secResultsDesignPrinciplesAndPatterns}.

\begin{table}[htp]
\centering
\caption{Age-related motivators, gamification mechanics, and effects on implementation}
\label{tab:age_mechanics_effects}
%\renewcommand{\arraystretch}{1.2}
%\begin{tabular}{@{}p{0.15\linewidth} p{0.23\linewidth} p{0.23\linewidth} p{0.23\linewidth}@{}}
\begin{tabular}{@{}%
  >{\raggedright\arraybackslash}p{0.15\linewidth}%
  >{\raggedright\arraybackslash}p{0.23\linewidth}%
  >{\raggedright\arraybackslash}p{0.23\linewidth}%
  >{\raggedright\arraybackslash}p{0.23\linewidth}@{}}

\toprule
\textbf{Age Group} & \textbf{Motivators} & \textbf{Gamification Mechanics} & \textbf{Effects on Implementation} \\
\midrule
Children & Immediate rewards; visual stimuli; collectibles & Animated avatars; sound effects; drag-and-drop & Short, reward-rich tasks \\
Adolescents & Competition; social recognition; leaderboards & Real-time feedback; quests; team points & Real-time quizzes; status comparisons \\
Young adults & Self-directed learning; sense of progress; autonomy & Cooperative mechanics; individualized tasks & Self-reflection via feedback histories \\
Adults & Goal orientation; relevance; actionable feedback & Self-paced learning; progress visualizations & Progress overviews; learning goals \\
Older adults & Clear feedback; experiences of success; social inclusion & Intuitive controls; clear explanations & Repetition; comprehensible structure \\
\bottomrule
\end{tabular}
\end{table}

\subsection{Risks of Misapplied Gamification}
While gamification can enhance motivation and engagement when designed with age-awareness, its misapplication may lead to adverse effects. Table \ref{tab:age_inhibitors_barriers} summarizes age-related inhibitors, barriers, and negative learning outcomes that occur when gamification is not aligned with developmental needs. These risks further highlight the importance of tailoring mechanics to learners' specific requirements.

\begin{table}[htp]
\centering
\caption{Age-related inhibitors, gamification mechanics, and effects on learning of misapplied gamification}
\label{tab:age_inhibitors_barriers}
%\begin{tabular}{@{}p{0.15\linewidth} p{0.23\linewidth} p{0.23\linewidth} p{0.23\linewidth}@{}}
\begin{tabular}{@{}%
  >{\raggedright\arraybackslash}p{0.15\linewidth}%
  >{\raggedright\arraybackslash}p{0.23\linewidth}%
  >{\raggedright\arraybackslash}p{0.23\linewidth}%
  >{\raggedright\arraybackslash}p{0.23\linewidth}@{}}
\toprule
\textbf{Age Group} & \textbf{Inhibitors} & \textbf{Gamification Mechanics} & \textbf{Effects on Learning} \\
\midrule
Children & Complex menus; unintelligible feedback & Long loading times; technical hurdles & Excessive attentional demands \\
Adolescents & Excessive performance pressure; constant points visibility & Opaque scoring systems & Frustration after failures \\
Young adults & Lack of goal transparency; irrelevant gamification & No adaptation to proficiency level & Unclear learning paths \\
Adults & Visual overload; forced interactions & Excessive gamification elements & Inflexible modules  \\
Older adults & Cluttered interfaces; icons too small & Overload due to too many options & Technology anxiety; insufficient accessibility  \\
\bottomrule
\end{tabular}
\end{table}

\section{Design Principles and Technical Patterns for Age-Specific Gamification}\label{secResultsDesignPrinciplesAndPatterns}
To answer RQ2, we derived actionable guidance on how gamification mechanics can be tailored to different age groups in multimedia learning environments. The results are presented as design principles (Section \ref{subsecDesignPrinciples}) and technical patterns for implementation (Section \ref{subsecTechnicalPatterns}).

\subsection{Design Principles}\label{subsecDesignPrinciples}
Based on our literature review, we abstracted five overarching principles that capture age-specific requirements for gamification in educational software: \textit{Feedback}, \textit{Progression}, \textit{Autonomy}, \textit{Coherence}, and \textit{Adaptivity}. Table \ref{tab:design-principles} gives an overview of the design principles and their functions. 

While these principles generally apply across the lifespan, their concrete realization varies by age group. The following analysis demonstrates how the same principles manifest differently across developmental stages.

\begin{table}[htbp]
  \centering
  \caption{Design principles and their functions}
  \label{tab:design-principles}
  \begin{tabular}{@{}ll@{}}
    \toprule
    \textbf{Design principle} & \textbf{Function} \\
    \midrule
    Feedback     & Learners receive direct feedback on their behavior. \\
    Progression  & Learning progress is made visible. \\
    Autonomy     & Users can make their own decisions. \\
    Coherence    & Content and game mechanics align in content and form. \\
    Adaptivity   & Content adapts to learners' abilities. \\
    \bottomrule
  \end{tabular}
\end{table}

\subsubsection{Feedback}
For children, feedback must be immediate and multisensory, relying on visual, auditory, or haptic cues to sustain attention \cite{becker_entwicklung_2022, hes_innovative_2024, hanus_assessing_2015}. Clear progress indicators and simple rewards (e.g., stars, avatars) provide orientation and motivation \cite{bodynek_gamification_2023}. For adolescents, it should be transparent and action-oriented to support self-evaluation, with clear task structures and direct feedback being particularly effective \cite{deci_selbstbestimmungstheorie_1993, tondello_framework_2017, noauthor_kim-studie_2024, wagenpfeil_gamification_2025}. Young adults benefit from personalized, functional feedback that emphasizes competence development, with differentiated feedback helping to sustain flow \cite{tondello_framework_2017, wesseloh_einsatz_2019, wagenpfeil_gamification_2025}. For adults, feedback is most useful when concise and differentiated, enabling monitoring of competence and informed decisions, while real-time feedback supports integration into work processes \cite{becker_einleitung_2022, widyakusuma_future_2024, beil_interface_2018, becker_konzept_2022}. Finally, older adults profit from simple, supportive, and accessible feedback, with read-aloud or audio support reducing cognitive load and facilitating persistence \cite{fietkau_quests_2019, basak_can_2008, technische_universitat_berlin_fakultat_fur_verkehrs-_und_maschinensysteme_gate_2015}.

\subsubsection{Progression}
For children, progression through short, playful steps and scaffolded challenges prevents overload \cite{hes_innovative_2024, becker_gamification_2022, bodynek_gamification_2023}. Adolescents respond well to competitive progression via leaderboards and quests, reinforced by multiplayer modes and group statistics that promote peer recognition \cite{tondello_framework_2017, faltermaier_entwicklungspsychologie_2014, wagenpfeil_gamification_2025}. Young adults value structured achievement systems and visualized milestones that make progression salient and maintain momentum \cite{wesseloh_einsatz_2019, wagenpfeil_gamification_2025}. For adults, structured milestones and clear dashboards communicate progression efficiently \cite{becker_einleitung_2022, widyakusuma_future_2024}. Older adults benefit from small, achievable steps that build confidence and sustain engagement \cite{fietkau_quests_2019, adeboye_docker_2025}.

\subsubsection{Autonomy}
For children, autonomy should be limited to guided choices as too much freedom can cause disorientation~\cite{tondello_framework_2017}. Adolescents require meaningful choices and opportunities for co-creation to support a growing need for self-determination \cite{tondello_framework_2017}. Flexible pacing and customization options that fit individual goals and contexts is important for young adults \cite{tondello_framework_2017, noauthor_kim-studie_2024}, while adults prefer flexibility within structured frameworks that enable self-regulated learning \cite{becker_einleitung_2022}. Finally, older adults need guided options and clear orientation to navigate tasks confidently \cite{fietkau_quests_2019}.

\subsubsection{Coherence}
For children, coherence emerges from storytelling and narrative embedding. Moreover, short, visually engaging microlearning units are especially effective \cite{becker_entwicklung_2022, hes_innovative_2024, bodynek_gamification_2023}. Adolescents benefit from authentic, socially relevant challenges that align with their lived contexts \cite{noauthor_kim-studie_2024}. Coherence is strongest for young adults when tasks are linked to professional development and real-world goals, with realism and meaningfulness amplifying engagement \cite{sailer_wirkung_2016, wagenpfeil_gamification_2025}. For adults, coherence increases when activities are directly relevant to professional and everyday contexts and when storytelling and modular content are embedded in workplace practice \cite{widyakusuma_future_2024, becker_konzept_2022}. For older adults, tasks should be closely tied to everyday usefulness \cite{basak_can_2008}.

\subsubsection{Adaptivity}
For children, adaptivity requires carefully adjusted difficulty to maintain engagement without frustration \cite{hes_innovative_2024, becker_gamification_2022}. Adolescents need a calibrated balance between challenge and skill to sustain flow \cite{tondello_framework_2017, noauthor_kim-studie_2024}. Young adults benefit from difficulty scaling with skill levels alongside adaptive content, modular learning paths, and customizable interfaces \cite{tondello_framework_2017, noauthor_kim-studie_2024}. Adults require adaptivity that tailors challenges to professional requirements and time constraints \cite{wesseloh_einsatz_2019, beil_interface_2018}. For older adults, it is important to reduce complexity. Moreover, it should be possible to gradually adjust the tasks to their individual capabilities, with clear navigation and large interaction targets further supporting accessibility \cite{fietkau_quests_2019, basak_can_2008, adeboye_docker_2025, technische_universitat_berlin_fakultat_fur_verkehrs-_und_maschinensysteme_gate_2015}.

\subsection{Technical Patterns}\label{subsecTechnicalPatterns}
While the design principles provide conceptual guidance, their implementation in multimedia learning environments requires suitable technical solutions. From the analysis of existing systems and design challenges, three recurring technical patterns can be identified that support age-aware gamification in practice: \textit{Feedback and Reward Systems}, \textit{Real-Time Interactivity and Social Mechanisms}, and \textit{Scalability and Personalization}.

Each pattern links common implementation problems to solutions and consequences for different age groups.

\subsubsection{Feedback and Reward Systems}
Motivation is strongly dependent on feedback quality and timing. Effective gamification requires differentiated, context-sensitive feedback and appropriate rewards delivered immediately after user actions~\cite{beil_gamification_2018}. This presupposes a rule-based event logic that evaluates actions server-side, generates feedback automatically, and ensures low latency \cite{swacha_representation_2018}. Reward schemes should integrate elements of surprise to stimulate curiosity \cite{wagenpfeil_gamification_2025}. A robust infrastructure integrates visual, auditory, and haptic feedback adaptively without cognitive overload, while efficient rendering and asset management maintain fluid interactions~\cite{wagenpfeil_gamification_2025}.

Intuitive interactions such as drag-and-drop benefit from indicators, animations, and transitions that support orientation~\cite{friedman_drag-and-drop_2023}. Mechanics like competitions, leaderboards, or badges should be rule-linked to performance and, if necessary, suppressible to avoid overload. Progress bars, quest systems, and status displays often rely on WebSocket or API interfaces for real-time synchronization \cite{swacha_representation_2018}. Transparency emerges from modular progress structures and badges, while dynamic rankings and team statistics can be generated through filters and database queries \cite{wagenpfeil_gamification_2025}. For longer learning processes, milestones, checklists, and feedback histories structure engagement \cite{sailer_wirkung_2016}. To remain motivating, feedback should be subtle, personalizable, and persisted server-side so that users can adjust preferences~\cite{wagenpfeil_gamification_2025}.

\subsubsection{Real-Time Interactivity and Social Mechanisms}
Collaborative and competitive learning contexts depend on low-latency, synchronous data transfer, and modular system architecture \cite{wagenpfeil_gamification_2025, beil_gamification_2018}. Children benefit from simple reaction games, animated avatars, and clear visual feedback \cite{sailer_wirkung_2016}. Adolescents respond well to competitive elements such as live leaderboards, time pressure, and points, which require stable server infrastructures with synchronized interfaces, load balancing, and container orchestration \cite{swacha_representation_2018, adeboye_docker_2025}. Young adults expect team functions, discussion spaces, and real-time communication, ideally built on modular backends with high throughput and caching \cite{wagenpfeil_gamification_2025}. Adults emphasize efficient interactions that align with workflow, including role-based access control, logging, and event triggers \cite{becker_positive_2024, wagenpfeil_gamification_2025}. Platforms like Moodle illustrate how collaborative features and status visualization are applied in professional contexts \cite{wagenpfeil_gamification_2025}. Older adults require clear, intuitive interfaces that support spontaneous reactions and social participation. For reliable communication, protocols such as WebRTC, MQTT, or Socket.IO are useful, complemented by safeguards like rate limiting, session management, and fallback strategies \cite{p_optimizing_2024, kumaran_secure_2024, pedreira_architecture_2020}.

\subsubsection{Scalability and Personalization}
Age-aware gamification depends on scalable, adaptive infrastructures. Modular designs, service-oriented interfaces, and cloud-based architectures provide the flexibility needed for large diverse learner groups \cite{beil_gamification_2018}. For children, simple interaction concepts, clear progressions, and age-appropriate paths are crucial, as demonstrated by products such as ABCmouse or Endless Alphabet \cite{wagenpfeil_gamification_2025}. Adolescents require dynamic learning paths, adaptive difficulty, and quest systems orchestrated server-side \cite{wagenpfeil_gamification_2025}. Young adults benefit from AI-based personalization systems such as ALEKS or DreamBox, which adjust content in real time and scale via cloud infrastructures \cite{wagenpfeil_gamification_2025, wang_cghit_2024}. Adults require integration with learning management systems, role management, and microservices to ensure smooth adoption \cite{wagenpfeil_gamification_2025}. Older adults profit from accessible interfaces with high contrast, simplified menus, and audio feedback \cite{technische_universitat_berlin_fakultat_fur_verkehrs-_und_maschinensysteme_gate_2015}.

\section{Discussion}\label{secDiscussion}
The results of this study highlight that gamification in educational contexts must be designed with age awareness in mind. Our mapping (Section \ref{secResultsAgeRelatedRequirements}) demonstrates that requirements differ substantially across developmental stages. %The derived design principles and technical patterns (Section \ref{secResultsDesignPrinciplesAndPatterns}) provide structured guidance for tailoring gamification mechanics to diverse learner groups.

The findings from Section \ref{subsecDesignPrinciples} show that the same five principles manifest differently across developmental stages. Thus, age-aware gamification does not depend entirely on new mechanics but on the differentiated application of universal principles that align with learners' motivational needs and cognitive abilities.

The technical patterns described in Section \ref{subsecTechnicalPatterns} complement the design principles by showing how conceptual requirements can be operationalized in practice. Feedback systems sustain motivation through low-latency reinforcement, real-time interactivity enables collaborative and competitive engagement, and scalable personalization architectures adapt learning experiences to the requirements of different age groups.

However, our investigation also showed that the benefits of gamification are accompanied by age-specific risks. Children are susceptible to overstimulation, where excessive cues reduce motivation and impede learning \cite{plass_foundations_2015}. Adolescents may experience performance pressure and dissatisfaction from constant comparison, and poorly designed gamification can even reduce motivation \cite{hanus_assessing_2015}. Young adults risk disengagement when superficial rewards are not tied to genuine learning progress, and weaker students may withdraw from participation \cite{becker_gamification_2022, barata_gamification_2015}. Adults may react negatively to overly playful visuals or distracting elements, and inappropriate mechanics can result in overload \cite{becker_einleitung_2022}. Older adults are particularly sensitive to complexity, and missing accessibility features can undermine trust and persistence \cite{technische_universitat_berlin_fakultat_fur_verkehrs-_und_maschinensysteme_gate_2015}.

In addition to pedagogical risks, the technical realization of gamification raises practical challenges. Ensuring low-latency responses, scalable personalization, and accessible interfaces requires robust infrastructures that may not be available in all educational settings. Cost, maintenance, and integration with existing learning management systems can further hinder adoption.

%\subsection{Ethical Challenges}
Because gamification deliberately targets motivation and emotions, its use also raises ethical challenges. Overreliance on extrinsic incentives can undermine intrinsic motivation \cite{deci_selbstbestimmungstheorie_1993, becker_konzept_2022}. Dark patterns, such as manipulative design strategies that influence behavior without user awareness, pose additional risks \cite{gray_dark_2018}. Examples include opaque leaderboards or delayed feedback loops that nudge learners toward compulsive engagement \cite{wagenpfeil_gamification_2025}.

Adaptive systems that rely on learning analytics and AI must also respect privacy and data protection standards, especially when applied to minors or vulnerable groups. Competitive elements can also create stress, exclusion, or negative social comparison, particularly among adolescents. To mitigate these risks, designers should ensure fairness and inclusivity, provide opt-outs for competitive mechanics, and tie incentives to meaningful learning outcomes \cite{stieglitz_gamification_2015, noauthor_jahresreport_2024}.

%\subsection{Summary}

\section{Conclusion \& Future Work}\label{secConclusionAndFutureWork}

In our paper, we investigate how gamification can be adapted to different age groups in multimedia learning environments. 
Based on the research questions we defined in Section~\ref{secintro}, we conducted a theory-led, integrative literature review. We identified age-related design requirements for gamification and developed insights into how gamification mechanics can be tailored to different age groups. 
As a result, we provided a mapping of age groups, gamification mechanics, and expected effects, which highlights how motivational needs and cognitive capacities differ across developmental stages. Moreover, we derived design principles that capture age-specific requirements and demonstrated how they manifest for children, adolescents, young adults, adults, and older adults. Finally, we formulated technical patterns that translate these principles into scalable, practical solutions for multimedia learning environments.

Our findings underline that gamification is not universally effective, but must be aligned with age-specific needs to foster engagement and avoid misaligned incentives. Age-aware design reduces the risk of overstimulation in children, excessive competition among adolescents, disengagement among young adults, lack of relevance for adults, and accessibility barriers for older learners.

At the same time, some limitations remain. Age is only one determinant of learner motivation, and factors such as cultural background, digital literacy, and personal preferences may shape responses to gamification. This limits the generalizability of age-based design guidelines and calls for further empirical validation. Moreover, technical feasibility and ethical concerns require ongoing attention, including privacy, fairness, and the prevention of manipulative design strategies.

In summary, while age-aware gamification holds significant potential for enhancing motivation and engagement in multimedia learning environments, its implementation requires careful attention to pedagogical, technical, and ethical boundaries. %Future research should therefore focus on empirical validation across diverse learner populations, long-term studies on learning outcomes, and systematic approaches to ensuring inclusivity and fairness.

The results from our study are design propositions that should be tested in future studies. Future work should therefore focus on empirical validation of the proposed mapping and principles across diverse learner populations. Longitudinal studies are needed to assess the sustained impact of age-aware gamification on learning outcomes. Further research should also examine how cultural, contextual, and individual differences interact with age in shaping motivational responses. Finally, the integration of emerging technologies such as adaptive AI tutors and immersive environments offers opportunities for extending age-aware gamification, but requires careful evaluation to ensure inclusivity, transparency, and ethical use.

\bibliographystyle{IEEEtran}
\bibliography{references}

\end{document}